\documentclass[conference]{IEEEtran}
\IEEEoverridecommandlockouts
\usepackage{cite}
\usepackage{amsmath,amssymb,amsfonts}
\usepackage{algorithmic}
\usepackage{algorithm}
\usepackage{graphicx}
\usepackage{subfigure}
\usepackage{textcomp}
\usepackage{xcolor}
\usepackage{multirow}
\usepackage{booktabs}
\usepackage{caption}
\usepackage{subcaption}
\makeatletter
\renewcommand{\maketag@@@}[1]{\hbox{\m@th\normalsize\normalfont#1}}
\makeatother
\vspace{-0.2cm}
\begin{document}
\title{Cooperative Semantic Knowledge Base Update Policy for Multiple Semantic Communication Pairs
}
\author{
	\IEEEauthorblockN{
		Shuling Li, 
		Yaping Sun, 
		Jinbei Zhang, 
		Kechao Cai,
            Hao Chen,
		Shuguang Cui and Xiaodong Xu}\vspace{-0.9em}
\thanks{S. Li, J. Zhang and K. Cai are with the Sun Yat-sen University, Shenzhen 518107, China. (email: lishling8@mail2.sysu.edu.cn, \{zhjinbei, caikch3\}@mail.sysu.edu.cn)}
\thanks{Y. Sun and H. Chen are with the Department of Broadband Communication, Pengcheng Laboratory, Shenzhen 518000, China. Y. Sun is also with the Shenzhen Future Network of Intelligence Institute (FNii-Shenzhen), Chinese University of Hong Kong (Shenzhen), Shenzhen 518172, China. (email: \{sunyp, chenh03\}@pcl.ac.cn)}
\thanks{S. Cui is with the School of Science and Engineering (SSE), the Shenzhen Future Network of Intelligence Institute (FNii-Shenzhen), and the Guangdong Provincial Key Laboratory of Future Networks of Intelligence, Chinese University of Hong Kong (Shenzhen), Shenzhen 518172, China. (email: shuguangcui@cuhk.edu.cn)}
\thanks{X. Xu is with the Beijing University of Posts and Telecommunications, Beijing 100876, China, and affiliated with the Department of Broadband Communication, Pengcheng Laboratory, Shenzhen 518000, China. (email: xuxiaodong@bupt.edu.cn)}
} 
\maketitle
\begin{abstract}

Semantic communication has emerged as a promising communication paradigm and there have been extensive research focusing on its applications in the increasingly prevalent multi-user scenarios. 
However, the knowledge discrepancy among multiple users may lead to considerable disparities in their performance. To address this challenge, this paper proposes a novel multi-pair cooperative semantic knowledge base (SKB) update policy. 
Specifically, for each pair endowed with SKB-enabled semantic communication, its well-understood knowledge in the local SKB is selected out and uploaded to the server to establish a global SKB, via a score-based knowledge selection scheme. The knowledge selection scheme achieves a balance between the uplink transmission overhead and the completeness of the global SKB. 
Then, with the assistance of the global SKB, each pair’s local SKB is refined and their performance is improved. 
Numerical results show that the proposed cooperative SKB update policy obtains significant performance gains with minimal transmission overhead, especially for the initially poor-performing pairs.

\end{abstract}

\section{Introduction}

Semantic communication is a revolutionary communication paradigm and has great potential to support intelligent connection \cite{letaiefEdgeArtificialIntelligence2022}. Different from bit-oriented traditional communication, semantic communication aims to transmit task-relevant information at the semantic level \cite{9679803, cddm}. With the support of the semantic knowledge base (SKB), extraneous details in the source message are filtered out, and only the essential information relevant to the specific task is transmitted. Such a paradigm signiﬁcantly reduces the transmission overhead and improves the accuracy of communication \cite{uav_swarm}. 

With the increasing popularity of emerging applications, such as smart transportation, mixed reality and e-health, a huge amount of devices are connected and multi-user scenarios have become widespread \cite{network_slicing,splitoverwirenet}. There have been some studies concentrating on multi-user semantic communication \cite{zhangMultiUser2022,Shaomultidevice}. Meanwhile, some works \cite{zero_shot_sun,moc-rvq} investigate the effective construction of SKB to improve semantic transmission efficiency.
However, existing works have not considered the knowledge discrepancy among multiple users.
Considering a scenario with a server and multiple pairs of end users. Within each pair, the transmitter and receiver share a common SKB, and communicate via semantic communication with the support of the SKB. Such a semantic communication pair is denoted as SemCom pair in this paper. It is noteworthy that each SemCom pair owns one local SKB and the local SKBs of different SemCom pairs may differ. 
In this case, different SemCom pairs may exhibit varying levels of performance on the same task, with some doing well and others not so well. 
The underperforming SemCom pairs may struggle to provide satisfactory quality of experience (QoE) and accomplish tasks efficiently.
Therefore, it is worth considering how to refine the SKB and improve the performance of poor-performing SemCom pairs in such situations.

In the aforementioned scenario, individual SemCom pairs possess diverse strengths and expertise, which are shown in their local SKBs.
Hence, inspired by the complementary strengths of multiple SemCom pairs, updating SKB via multi-pair cooperation is a feasible strategy.  
Existing works \cite{dynamic_channel,FL_audio} mainly combine federate learning (FL) with semantic communication as a solution for multi-user cooperation, where the FL technology has powerful learning ability and facilitates effective model training and aggregation \cite{9530714}. 
Each local model is trained with local data independently, whose model parameters are periodically transmitted to the server for aggregating as a global model. By updating each local model to the aggregated global model, the performance of each user is improved effectively. 

However, FL-based methods necessitate periodical transmission of model parameters, which may not be friendly for large-scale neural networks and dynamic channel environments. In addition, the aforementioned works mainly focus on the the update of local model, while ignoring the update of local SKB. It should be noted that the mismatch between the local model and local SKB may yield poor performance. 
For instance, for a local SKB that contains ambiguous information about some categories, if the local model is updated but the local SKB is not updated accordingly, the precision of related tasks, such as image retrieval task and image classification task, will be impaired. Thus, the update of SKB is of great significance to a semantic communication system. 

To tackle the above issues, this paper proposes a novel multi-pair cooperative SKB update policy, by which each local SKB is updated with less ambiguity and mistakes, and each SemCom pair acquires advancement in performance.
The main contributions of this paper are summarized as follows.
\begin{itemize}
        \item A cooperative SKB update policy for multiple SemCom pairs is proposed, in which the well-comprehended knowledge from different local SKBs is shared for SKB update. With the updated SKB via cooperation, significant performance improvement is observed for the data initially misunderstood. 
    \item A score-based knowledge selection scheme is designed to select the uploaded knowledge and flexibly adjust the uplink transmission overhead. The uplink transmission overhead of each SemCom pair is adjustable by a pre-configured threshold, which enables sharing comprehensive knowledge with minimal transmission overhead.
    \item Numerical results show that the SemCom pairs exhibit promising performance gains after cooperation, which verifies the effectiveness of the proposed cooperative SKB update policy.

\end{itemize}

\section{System Model}
\begin{figure}[t]
	\centering
	\includegraphics[width=0.9\linewidth]{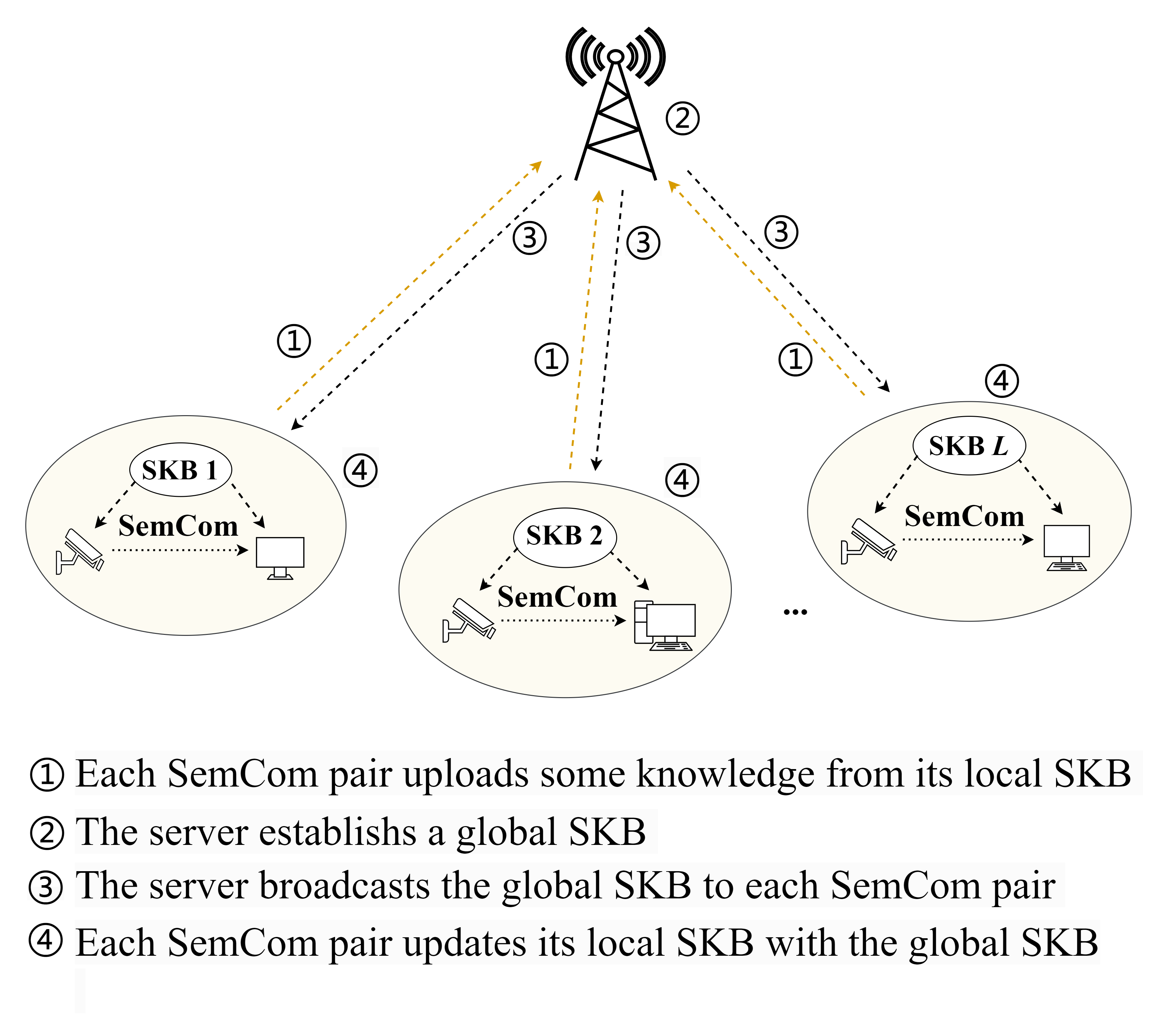}
	\caption{System model and cooperative SKB update process. } 
	\label{1-SKB_model}
\end{figure}

As shown in Fig. \ref{1-SKB_model}, we consider a multi-user system which consists of a server and $L$ SemCom pairs. Each SemCom pair contains a transmitter, a receiver and a shared SKB, maintaining an end-to-end semantic communication model. 
In this system, all SemCom pairs collaborate to update their respective local SKBs, which can bring performance enhancement for the SemCom pair with misunderstanding. 

\subsection{Semantic knowledge base (SKB)}
As the shared background knowledge between both ends within one SemCom pair, SKB represents the SemCom pair's knowledge level. Similar to \cite{skb_gsc}, class-level attribute vectors are utilized to construct the SKB in this paper, where each attribute vector associates with a particular category and reflects the states of specific attributes.

In particular, there are two features of the SKB in this paper, which are different from \cite{skb_gsc}. First, each SemCom pair's SKB may vary due to the difference in knowledge levels. Take image classification task as an example, some SemCom pairs can recognize excavators while other SemCom pairs are unable to accurately identify excavators and forklifts.
Second, for each SemCom pair, with the update of their local model, their local SKB will be updated correspondingly, which ensures the consistency between the local model and local SKB.
Let $\mathcal{M}\overset{\Delta}{=}\{1,2,\dots,M\}$ denote the set of classes, $d$ denote the number of attributes and $\boldsymbol{k}_{m} \in\mathbb{R}^{d}$ denote the attribute vector of the $m$-th class, respectively. Then, let $\mathcal{K}\overset{\Delta}{=}\{\boldsymbol{k}_{m}\in\mathbb{R}^{d} \}_{m\in \mathcal{M}}$ denote the set of attribute vectors, i.e., the SKB.

\subsection{SKB-enabled semantic communication model}
The framework of the semantic communication model within each SemCom pair is same as the setup of \cite{skb_gsc}, which supports both image classification task and image generation task. At the transmitter, a semantic encoder extracts semantic information $\boldsymbol{s} \in \mathbb{R}^{d}$, which is regarded as the predicted attribute values of the input image. After finding out the vector that is most similar to $\boldsymbol{s}$ in the local SKB, $\boldsymbol{s}'$, its index $v$ in the SKB is transmitted to the receiver. 
Thus, the transmission load is greatly reduced by sending the index $v$ rather than $\boldsymbol{s}$. Assume the index $v$ as the class label, thus the classification task is executed at the sender.
In addition, a generative encoder is equipped at the transmitter, whose input includes the source image and the selected attribute vector $\boldsymbol{s}'$.  

The image generation task has two modes. When the testing compression ratio (CR, defined as the ratio between the number of transmitted symbols and input symbols) $B$ is less than training CR $\theta$, only relevant indices are transmitted to generate images that are of the same classes as the source images. Otherwise, when $B\geq \theta$, the output of the generative encoder $\boldsymbol{z}$ is transmitted with the indices for image reconstruction, where the indices are transmitted without error. The physical channel is set as additive white Gaussian noise (AWGN) channel when transmitting $\boldsymbol{z}$, and the received signal is denoted as $\hat{\boldsymbol{z}}=\boldsymbol{z}+\boldsymbol{n}$, where $\boldsymbol{n} \sim N(0,\sigma^2I)$. At the receiver, the result of image classification task is obtained by index $v$. 
When $B<\theta$, the generated class-consistent image is acquired by processing $s'$ and random Gaussian samples $\boldsymbol{z}_p$ with a generative decoder. When $B\geq \theta$, the source image is reconstructed by decoding $\boldsymbol{s}'$ and $\hat{\boldsymbol{z}}$.

Note that the SKB is crucial to the semantic communication model and plays a significant role in the whole performance. The communication effectiveness is diminished if the SKB is perplexing or its update is undesirable, and these inherent limitations are difficult to overcome by single SemCom pair. Thus, it is necessary to modify the SKB via multi-pair collaboration.

\subsection{Performance metric}
Since the SKB indicates class-level information, the evaluation metric is supposed to reflect the comprehension of each category. $F_1$-score \cite{Goutte2005API} is adopted as the performance metric in this paper. Compared with traditional classification accuracy focusing on the whole accuracy of all categories, $F_1$-score emphasizes the accuracy of each category.
Calculating the $F_1$-score for each class can be viewed as a binary classification problem. Given a specific class, all samples can be divided into belonging to this class (Positive) or not (Negative). The predicted results can be divided into correctly predicting (True) or incorrectly (False). 
All four possible cases are shown in Table \ref{conf_mat}, which is named as confusion matrix.
Then, precision and recall can be computed as follows, respectively.

\begin{footnotesize}
\begin{equation}
Precision = \frac{TP}{TP+FP},
\end{equation}
\end{footnotesize}
\begin{footnotesize}
\begin{equation}
Recall = \frac{TP}{TP+FN}.
\end{equation}
\end{footnotesize}

\noindent{The $F_1$-score is calculated as the harmonic average of precision and recall, which can be written as }

\begin{footnotesize}
\begin{equation}\label{eq:f1}
F_1\mbox{-}score = \frac{2 \times Precision \times Recall}{Precision+Recall}.
\end{equation}
\end{footnotesize}

\noindent{where $F_1$-score $\in[0,1]$. The $F_1$-score provides a balance between precision and recall, making it a comprehensive metric for evaluating classification performance.} 
\begin{table}[ht]
 \renewcommand{\arraystretch}{1.5}
\centering
\caption{Confusion matrix.} 
\scalebox{0.9}{
\begin{tabular}{|cc|cc|l}
\cline{1-4}
\multicolumn{2}{|c|}{\multirow{2}{*}{}}                        & \multicolumn{2}{c|}{Predicted label}     &  \\ \cline{3-4}
\multicolumn{2}{|c|}{}                                         & \multicolumn{1}{c|}{Positive} & Negative &  \\ \cline{1-4}
\multicolumn{1}{|c|}{\multirow{2}{*}{Actual label}} & Positive & \multicolumn{1}{c|}{ True Positive (TP)}       &  False Negative (FN)       &  \\ \cline{2-4}
\multicolumn{1}{|c|}{}                              & Negative & \multicolumn{1}{c|}{ False Positive (FP)
}       & True Negative (TN)       &  \\ \cline{1-4}
\end{tabular}
}\label{conf_mat}
\end{table}

\section{Multi-pair Cooperative SKB Update Policy}
In this section, we illustrate the process of the multi-pair cooperative SKB update policy, and some details about the update of local SKB and knowledge selection scheme in uploading phase.
\subsection{Update of local SKB}
For each SemCom pair, with the update of the local model, the local SKB needs to be updated correspondingly. Since the semantic encoder is able to predict the attribute values of the input image, the output of the semantic encoder $\boldsymbol{s}$ can represent the SemCom pair's current knowledge level. Specifically, using a testing dataset with an equal sample number of each category as input data, some predicted attribute vectors are obtained by the semantic encoder. 
The average of the vectors which correspond to the $m$-th class is regarded as the updated knowledge of that category. 
Let $\mathcal{X}_m^{test} = \{x_{m,1}, x_{m,2},\dots,x_{m,N_m} \}$ denote the set of testing images belonging to the $m$-th class, where $m\in \mathcal{M}$ and $N_m$ denotes the number of images in the $m$-th class. The average attribute vector $\bar{\boldsymbol{s}}_m$ for the $m$-th class is calculated as $\bar{\boldsymbol{s}}_m = \frac{1}{N_m}\sum_{i=1}^{N_m} \boldsymbol{s}_{m,i}$.
Then, the updated SKB $\mathcal{K}^U$ can be written as
\begin{equation}
\mathcal{K}^U = \{\bar{\boldsymbol{s}}_m\}_{m\in \mathcal{M}}.
\end{equation}
The updated local SKB reflects the current comprehension of both the transmitter and receiver, which replaces the original local SKB in the testing phase and next training phase.

\begin{algorithm}[t]
	\renewcommand{\algorithmicrequire}{\textbf{Input:}}
	\renewcommand{\algorithmicensure}{\textbf{Output:}}
	\caption{The proposed cooperative SKB update policy}
	\label{alg:1}
    \begin{algorithmic}[1]
		\REQUIRE Local data $\mathcal{X}_l$,  Initial local SKB $\mathcal{K}^I_{l}$,  Testing data $\mathcal{X}^{test}$ for updating SKB 
                \FOR {each SemCom pair} 
                    \STATE Train local model with $\mathcal{X}_l$ and $\mathcal{K}^I_{l}$
                    \STATE Update local SKB as $\mathcal{K}^U_{l}$ (with $\mathcal{X}^{test}$ and $\mathcal{K}^I_{l}$)
                    \STATE Upload attribute vectors $\mathcal{S}_l$ via the score-based knowledge selection scheme
                \ENDFOR
                \STATE The server forms the global SKB $K^G$ and broadcasts $K^G$ to each SemCom pair
                \FOR {each SemCom pair}
                    \STATE Finetune the local model with $\mathcal{X}_l$ and $\mathcal{K}^G$
                    \STATE Update local SKB as $\mathcal{K}^E_{l}$ (with $\mathcal{X}^{test}$ and $\mathcal{K}^G$)
                \ENDFOR
    \end{algorithmic}  
\end{algorithm}

\begin{algorithm}[t]
	\renewcommand{\algorithmicrequire}{\textbf{Input:}}
	\renewcommand{\algorithmicensure}{\textbf{Output:}}
	\caption{The score-based knowledge selection scheme}
	\label{alg:2}
    \begin{algorithmic}[1]
		\REQUIRE  $\mathcal{X}^{test} = \{ \mathcal{X}_m^{test} \}_{m\in \mathcal{M}}$, $\gamma$
                \FOR {each SemCom pair} 
                    \FOR {$m \text{ in }\mathcal{M}$}
                    \STATE Calculate the $F_1$-score $F_1^{l,m}$ by (\ref{eq:f1})
                    \IF{$F_1^{l,m}>\gamma$} 
                    \STATE Upload $F_1^{l,m}$ and $\bar{\boldsymbol{s}}_{l,m}$ to the server
                    \STATE Add $\bar{\boldsymbol{s}}_{l,m}$ into $\mathcal{S}_l$
                    \ENDIF
                    \ENDFOR
                \ENDFOR
    \end{algorithmic}  
\end{algorithm}

\addtolength{\topmargin}{0.15cm}
\subsection{Cooperative SKB update policy}

After training the local model and updating the local SKB, the updated local SKB is related to the original local SKB, limiting the modification of SKB. If the original local SKB is misguided, the updated local SKB is hard to be rectified only by the initial SKB and may still contain some misunderstandings. 
To cope with this issue, each SemCom pair can collaborate to make up for each other's deficiencies, as shown in Fig. \ref{1-SKB_model}.  

First, each SemCom pair computes the $F_1$-score of each class during updating the local SKB. Based on a score-based knowledge selection scheme, they select out their superior knowledge about some categories. 
Let $\mathcal{M}_l$ and $\mathcal{S}_l=\{\bar{\boldsymbol{s}}_{l,m'}\}_{m'\in \mathcal{M}_l}$ denote the set of selected categories and selected knowledge from the $l$-th SemCom pair, respectively. And let $F_1^{l,m'}$ denote the class-level $F_1$-score for the $l$-th SemCom pair. Then, the chosen knowledge $\bar{\boldsymbol{s}}_{l,m'}$ and corresponding $F_1$-score $F_1^{l,m}$ are uploaded to the server by the $l$-th SemCom pair. 
We assume that the transmitter in each SemCom pair is closer to the server than the receiver. Considering limited uplink energy, it is feasible to choose the transmitter to execute the upload operation. 

At the server, the attribute vectors with the maximum $F_1$-score of each category are aggregated to form global SKB $\mathcal{K}^G$. The global SKB $\mathcal{K}^G$ combines the best understanding of each category from all SemCom pairs, specifically referring to the current stage. Next, the server broadcasts the global SKB $\mathcal{K}^G$ to each SemCom pair for finetuning the local model and updating the local SKB. 
Let $\mathcal{K}^U_l$ denote the updated local SKB before cooperation and $\mathcal{K}^E_l$ denote the updated local SKB which is enhanced by $\mathcal{K}^G$ after cooperation, where the subscript denotes the $l$-th SemCom pair. Owing to the collaboration, $\mathcal{K}^E_l$ contains more sensible comprehension and less ambiguity, in contrast to $\mathcal{K}^U_l$. The cooperation process is shown in Algorithm \ref{alg:1}.

\subsection{Score-based knowledge selection scheme}
In order to adjust the uplink transmission overhead, we design a score-based knowledge selection scheme, as shown in Algorithm \ref{alg:2}. When updating the local SKB with specific testing data, the $F_1$-score of each class $F_1^{l,m}$ is calculated. Then, $F_1^{l,m}$ is compared with a predetermined threshold $\gamma$, where $\gamma \in [0,1]$. If $F_!^{l,m}$ is larger than $\gamma$, the corresponding attribute vector $\bar{\boldsymbol{s}}_{l,m}$ in the local SKB $\mathcal{K}^U_l$ is believed as excellent knowledge of the $l$-th SemCom pair, which needs to be uploaded together with $F_1^{l,m}$ to the server. Otherwise, if $F_1^{l,m}\leq \gamma$,  $\bar{\boldsymbol{s}}_{l,m}$ is not good enough to be uploaded. 

Therefore, the threshold $\gamma$ controls the tradeoff between the uplink transmission overhead and the completeness of the global SKB $\mathcal{K}^G$.
As $\gamma$ grows to one, the criterion of uploading knowledge will be more strict, and less attribute vectors in the local SKBs will be uploaded, leading to fewer transmission overhead. However, for  certain categories, their attribute vectors may not be uploaded due to not meeting the prerequisite, causing a lack of information in these categories in the global SKB.
Conversely, when $\gamma$ is going down to zero, the comprehensiveness of the global SKB is ensured while the uplink transmission overhead increases.

Let $M_l$ denote the length of $\mathcal{M}_l$, $Q=10$ denote the quantization level and $R_l$ denote the uplink transmission rate of the $l$-th SemCom pair, the uplink transmission latency of the $l$-th SemCom pair is given by $T_l = \frac{QM_l(d+1)}{R_l}$. 
The path loss $g_l$ between the $l$-th SemCom pair and the server is formulated as $g_l = \beta_0(\frac{D_l}{D_0})^{-\zeta}$, where $\beta_0=-$30 dB denotes the path loss at the reference distance $D_0=$ 10 m, $D_l$ denotes the distance between them, and $\zeta=3$ denotes the path loss exponent. Let $B_U=$ 1 MHz denote the allocated uplink bandwidth, $P=$ 10 dBm denote the uplink power of each SemCom pair, and $N_0=-$60 dBm denote the noise power. Then, the uplink transmission rate of the $l$-th SemCom pair is given by $R_l=B_U\log_{2}{(1+\frac{Pg_l}{B_UN_0})}$.

\begin{figure}[t]
	\centering
	\includegraphics[width=0.89\linewidth]{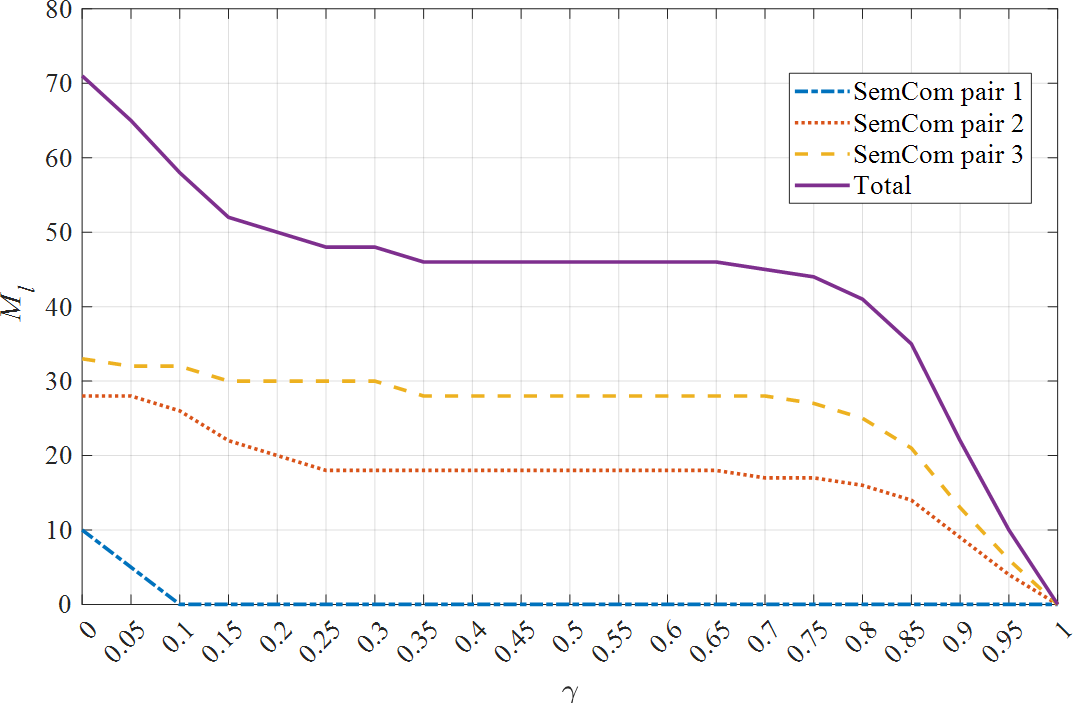}
	\caption{$M_l$ versus $\gamma$.}
	\label{m_vs_r}
\end{figure}

\begin{figure}[t]
	\centering
	\includegraphics[width=0.89\linewidth]{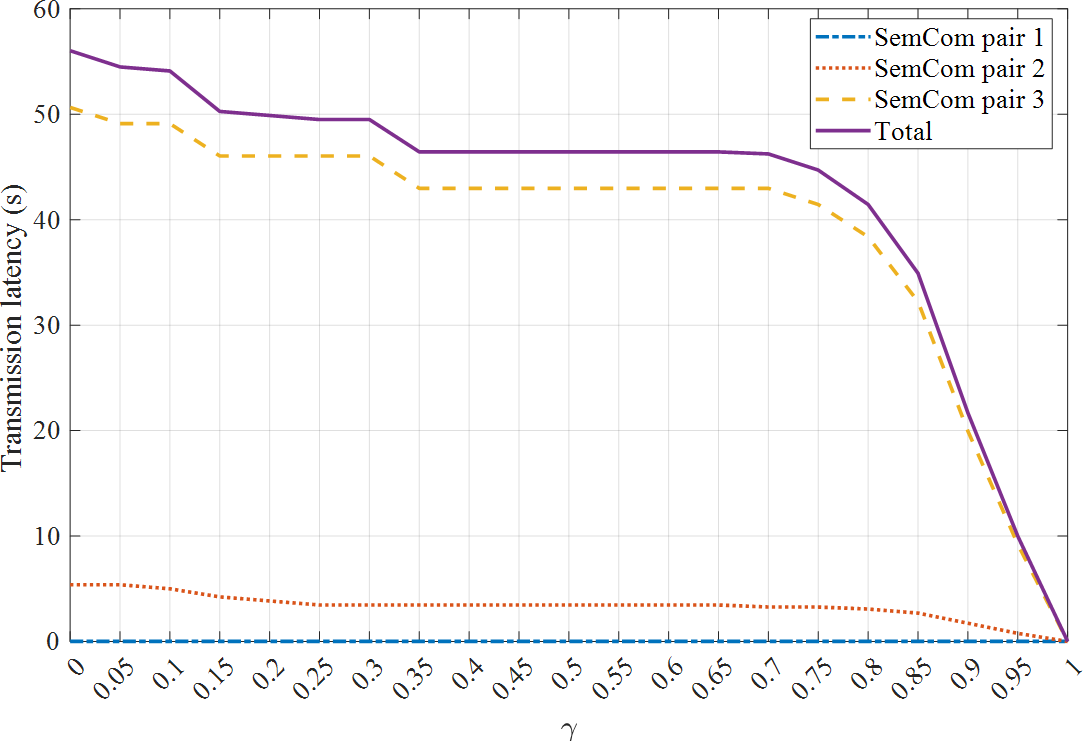}
	\caption{Transmission latency versus $\gamma$.}
	\label{t_vs_r}
\end{figure}

\section{Numerical Results}

\subsection{Simulation settings}
The adopted dataset is the vehicle subset (denoted as LAD\_vehicles) from Large-scale Attribute Dataset (LAD) \cite{lad_dataset}, consisting of 33 classes with sample sizes exceeding 300. Each category is annotated with 81 attributes, so $d=81$ and the size of SKB is $33\times81$. Take 80 samples from each category as the shared testing dataset for updating the SKB. In the remaining data, randomly take 150 samples from each category as the local data for each SemCom pair. Thus, there exists some overlap among different local data. All images are pre-processed into size 256 × 256 and all attribute values are normalized to [0,1]. 

In this section, we mainly consider the case of training CR $\theta=0.055$ and testing CR $B<\theta$, where the image generation task with class-consistent generation mode is executed. The number of SemCom pairs is set as $L=3$. The distances between each SemCom pair and the server are set as $D_1=$ 50 m, $D_2=$ 150 m, $D_3=$ 300 m, respectively.
For SemCom pair 1, the elements of its initial local SKB $\mathcal{K}^I_1$ are all set as 0.5, indicating that SemCom pair 1 is unable to distinguish all categories. For SemCom pair 2, the attribute values of the first 15 classes in $\mathcal{K}^I_2$ are set as 0.5, while the attribute values of remaining classes are the default attribute table from the dataset  with Gaussian noise $\epsilon\sim N(0,0.64)$. This means SemCom pair 2 is confused with the first 15 classes but know something about other classes. For SemCom pair 3, the attribute values of the last 5 classes in $\mathcal{K}^I_3$ are set as 0.5, while the setup of remaining classes is same as the latter 18 classes in $\mathcal{K}^I_2$. This indicates SemCom pair 3 is puzzled about the last 5 classes but know something about other classes.

\begin{figure}[H]
    \centering
    \subfigure[SemCom pair 1]{
        \includegraphics[width=0.7\linewidth]{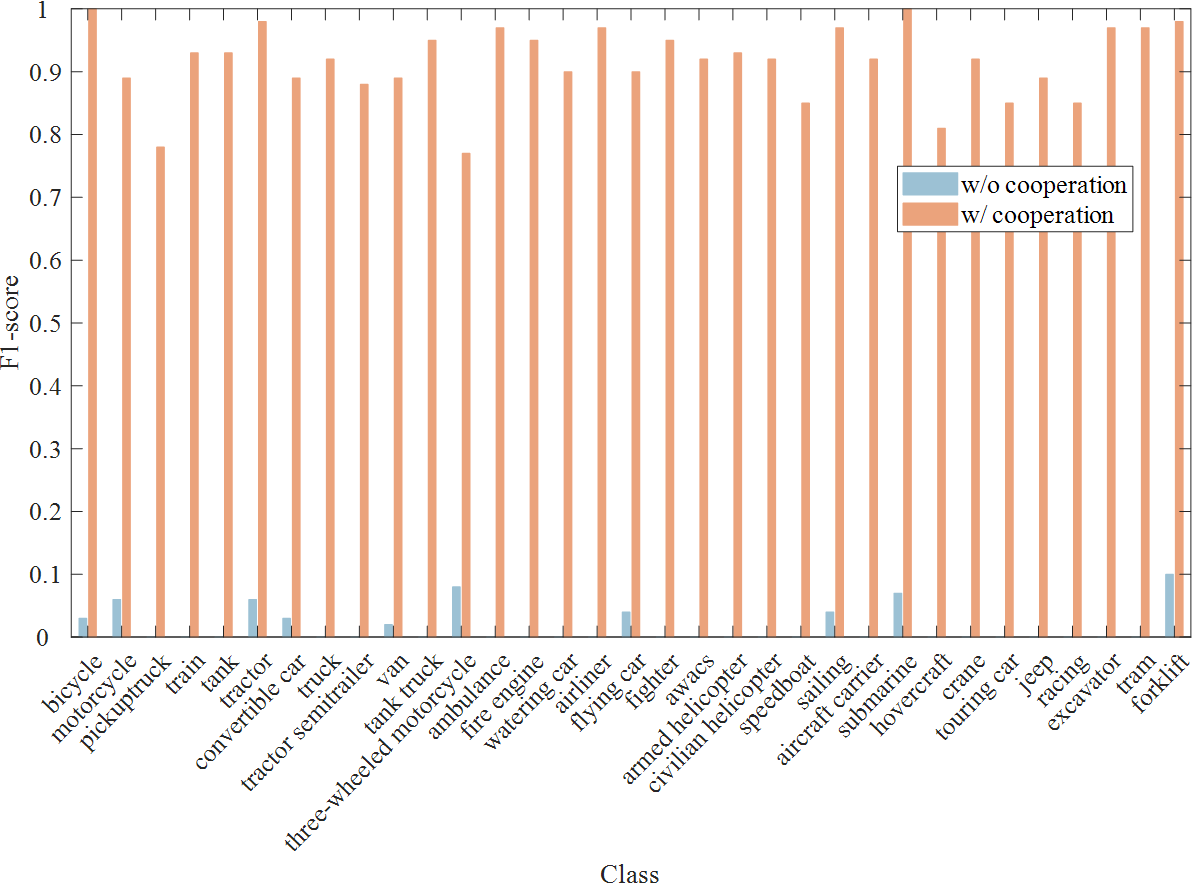}
    }
    \subfigure[SemCom pair 2]{
	\includegraphics[width=0.85\linewidth]{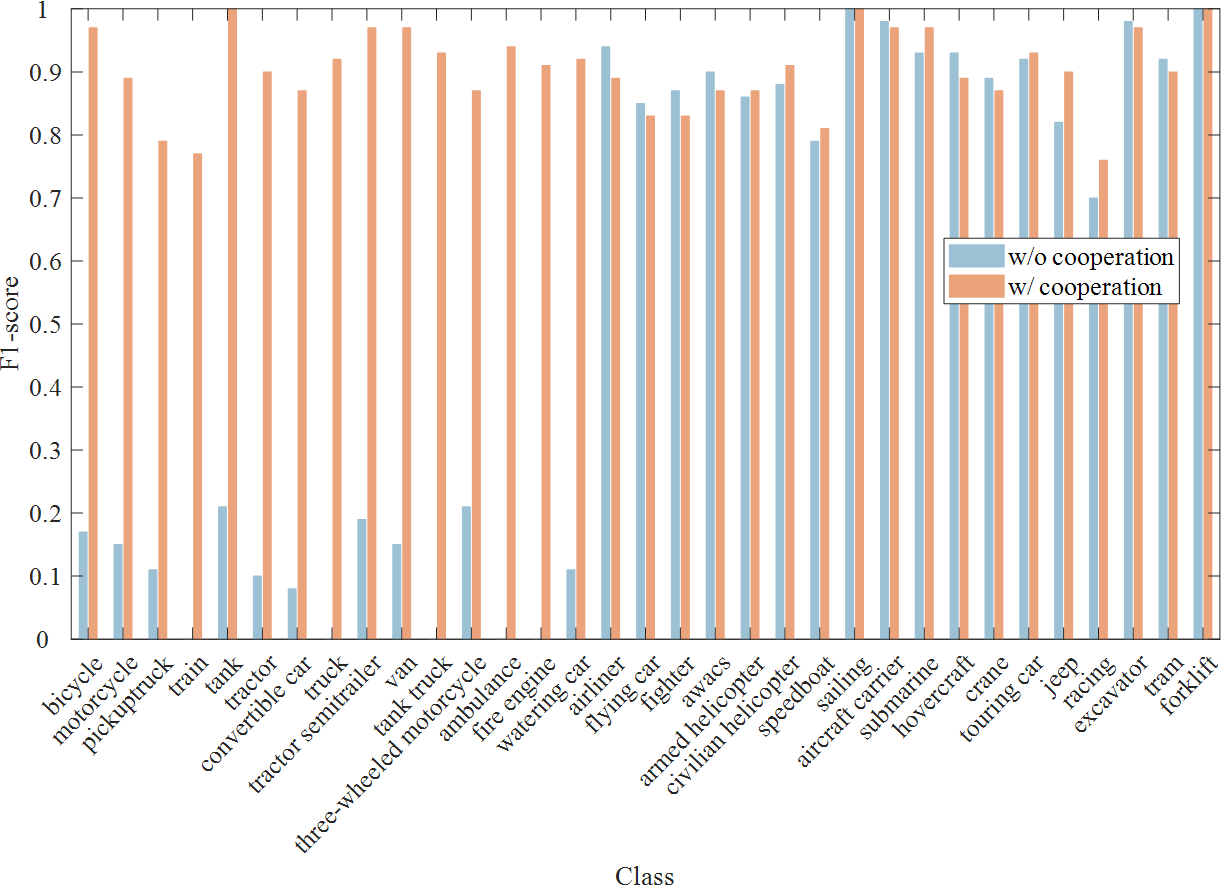}}
     \subfigure[SemCom pair 3]{
	\includegraphics[width=0.85\linewidth]{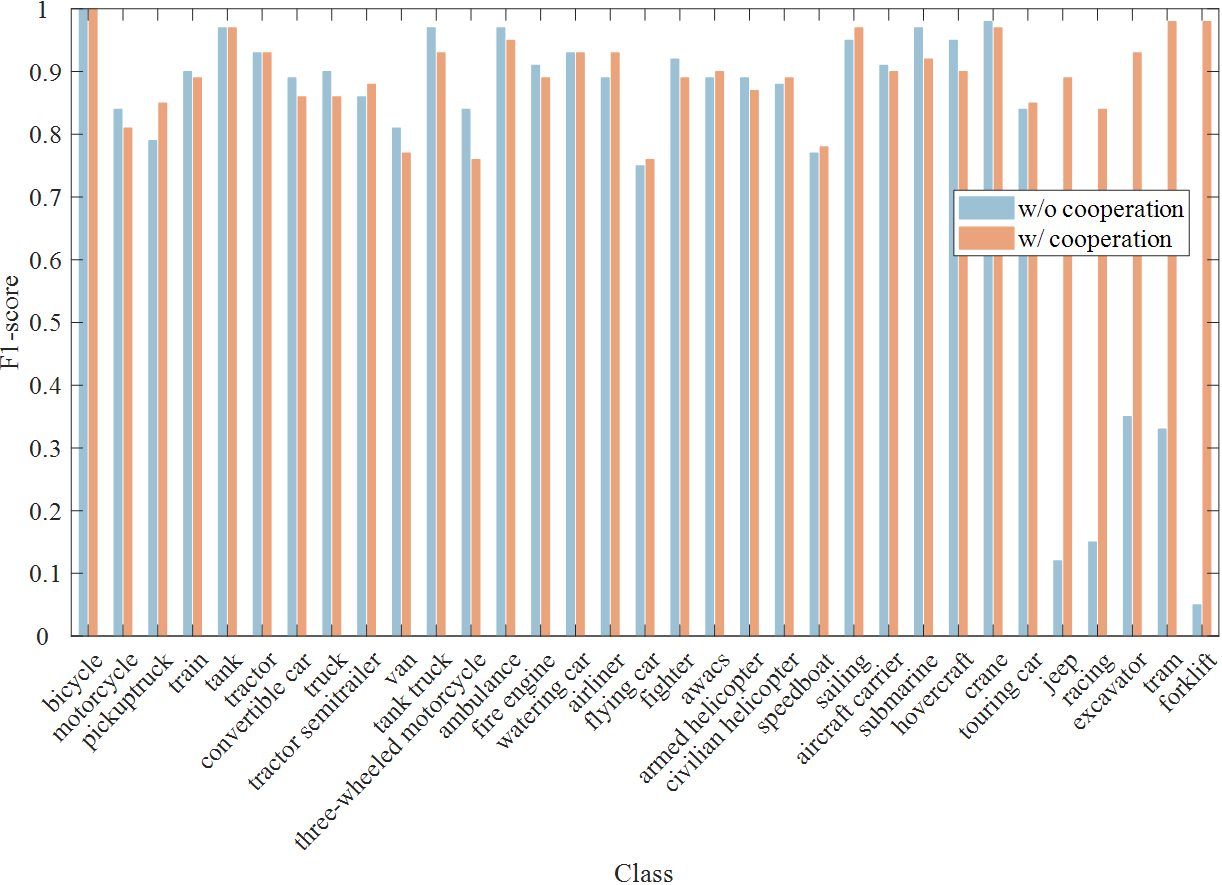}}
    \caption{$F_1$-scores of each class. Blue bars and orange bars represent the $F_1$-scores without cooperation and with cooperation, respectively.}
    \label{classification task}
\end{figure}

For image classification task, $F_1$-score of each class and Macro $F_1$-score are exploited as the metrics. Macro $F_1$-score represents the average $F_1$-score of all classes, which is given by Macro $F_1^l$ $=\frac{\sum_{m=1}^{M} {F_1^{l,m}}}{M}$. For image generation task, we use Fr\'{e}chet Inception Distance (FID) score \cite{fid} to measure the distribution discrepancy between the generated images and real images. The lower FID score indicates the smaller difference, and better quality of the generated images.

\begin{figure*}[t]
	\centering
	\includegraphics[width=0.7\linewidth]{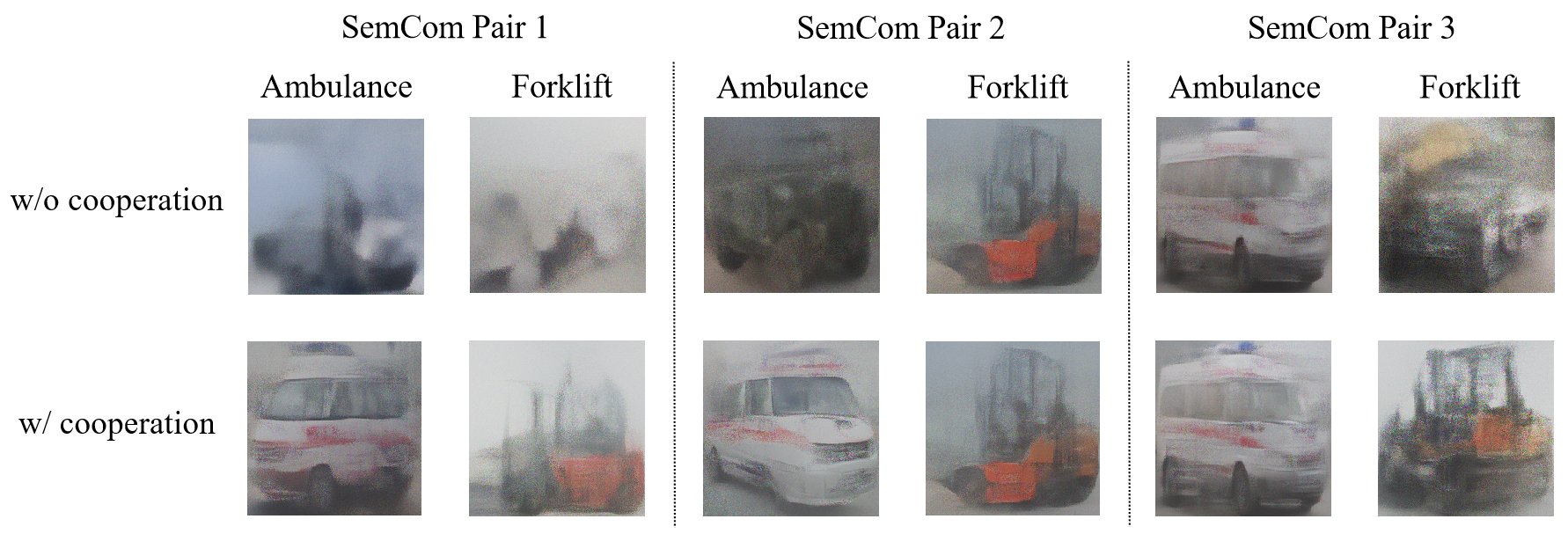}
	\caption{Visual comparisons of the generated images.}
	\label{generation task}
\end{figure*}

\addtolength{\topmargin}{-0.07cm}
\subsection{Simulation results}

Fig.~\ref{m_vs_r} shows the number of uploaded attribute vectors $M_l$ of each SemCom pair versus $\gamma$, where ``Total" denotes the sum of $M_l$, i.e., $M_1+M_2+M_3$. It can be observed that with the increase of $\gamma$, the transmission load is reduced, due to less qualified knowledge is selected for sharing.
When the threshold $\gamma$ grows, $M_3$ is always higher than $M_1$ and $M_2$, and $M_1=0$. This is because SemCom pair 1 is hard to distinguish the local data and its local SKB is unqualified. Besides, SemCom pair 3 knows more than SemCom pair 2, thus more information can be uploaded for SemCom pair 3. The total value of $M_l$ is below $M=$ 33 when $\gamma$ is larger than 0.85, which causes the incompleteness of the global SKB. Therefore, we choose $\gamma=0.85$ in the following simulation results, where $M_1=0$,  $M_2=14$ and $M_3=21$.

Fig. \ref{t_vs_r} depicts the transmission latency of each SemCom pair versus $\gamma$, where ``Total" denotes the total transmission latency of the three SemCom pairs. As the transmission latency is affected by the transmission load, the transmission latency gets lower with larger $\gamma$, which is in accordance with Fig. \ref{m_vs_r}. The transmission latency of SemCom pair 3 is obviously higher than that of SemCom pair 2, because of higher transmission load, i.e., $M_3>M_2$ and further distance to the server, i.e., $D_3 > D_2$. It can be seen that when $\gamma=0.85$, the total transmission latency is relatively lower, whose value is 34.92 s. And the transmission latency of the three SemCom pairs is 0 s, 2.69 s and 32.23 s, respectively.

Fig. \ref{classification task} compares the $F_1$-scores of each class and Table \ref{avg_f1}. shows the Macro $F_1$-scores for each SemCom pair. It can be observed that at first, each SemCom pair exhibits poor recognition performance for the categories with inappropriate understanding. However, through collaboration, each SemCom pair demonstrates significant performance improvements for these categories. 
For the categories that are originally well-understood, the $F_1$-scores remain similar performance after cooperation. The Macro $F_1$-scores for each SemCom pair rise notably after cooperation, especially for SemCom pair 1 and SemCom pair 2.

\addtolength{\topmargin}{0.095cm}
Some visual examples are shown in Fig. \ref{generation task}, where ambulance is the 13th class and forklift is the last class.
For SemCom pair 1, there are no distinct objects in the initial generated images due to its failure in differentiating all categories. Benefiting from the multi-pair cooperation, this SemCom pair is able to generate recognizable objects, e.g., ambulance and forklift. SemCom pair 2 first confounds categories such as jeeps and ambulances, but is able to distinguish forklifts. After cooperation, SemCom pair 2 gains the capability to generate ambulances and maintains the generation ability of forklifts. Similarly, SemCom pair 3 initially confuses forklifts and excavators, resulting in the generated forklifts being similar to excavators. With the assistance of the cooperative SKB update policy, SemCom pair 3 is able to generate forklifts accurately.
The comparison of FID scores is shown in Table \ref{gen_fid}. All FID scores of the three SemCom pairs get lower after cooperation, which indicates the generated images are more similar to the real images and verifies the effectiveness of the proposed multi-pair cooperation scheme.

\section{Conclusions}

In this paper, a cooperative SKB update policy for multiple SemCom pairs is proposed. Part of local SKB in each SemCom pair is selected by a score-based knowledge selection scheme for uploading to the server. Then, a global SKB is formed and shared to each SemCom pair. With the support of the global SKB, each SemCom pair refines their local SKB and enhances their performance. The knowledge selection scheme balances the tradeoff between the uplink transmission overhead and the completeness of the uploaded knowledge. 
Numerical results verify that the proposed scheme exhibits significant performance improvements,  especially for initially underperforming SemCom pairs.

\begin{table}[t]
\setlength{\abovecaptionskip}{0.2cm}
  \begin{center}
    \caption{Macro $F_1$-scores for each SemCom pair.}\label{avg_f1}
    \scalebox{0.9}{
    \begin{tabular}{c c c c} 
      \toprule
      &SemCom pair 1&SemCom pair 2&SemCom pair 3 \\
      \midrule
        w/o cooperation& 0.016& 0.535& 0.791\\
        w/ cooperation& 0.915& 0.902& 0.892\\
      \bottomrule
    \end{tabular}
    }
  \end{center}
  \end{table}

\begin{table}[t]
  \begin{center}
    \caption{FID scores of image generation task.}\label{gen_fid}
    \scalebox{0.9}{
    \begin{tabular}{c c c c} 
      \toprule
      &SemCom pair 1&SemCom pair 2&SemCom pair 3 \\
      \midrule
        w/o cooperation& 265.20	 & 264.44& 258.34 \\
        w/ cooperation& 247.83& 242.90& 238.93 \\
      \bottomrule
    \end{tabular}
    }
    \vspace{-0.4cm}
  \end{center}
\end{table}

\section*{Acknowledgement}
This work was supported in part by National Key R\&D Program of China under Grant 2022YFB2902700, NSF China (Grant No. 62471505, 62301471, 62293482, 62071501, 62202508), Shenzhen Science and Technology Program (Grant JCYJ20220818102011023, 20220817094427001, ZDSYS20210623091807023), and the Major Key Project of PCL Department of Broadband Communication (PCL2023AS1-1).

\bibliographystyle{ieeetr}
\bibliography{main}

\end{document}